\documentstyle[12pt,fleqn]{article}
\def\la{\mathrel{\mathpalette\fun <}}

\def\fun#1#2{\lower3.6pt\vbox{\baselineskip0pt\lineskip.9pt
         
\ialign{$\mathsurround=0pt#1\hfill##\hfil$\crcr#2\crcr\sim\crcr}}}
\begin{document}
\begin{titlepage}
\null\vspace{-72pt}
\begin{flushright}
{\footnotesize
FERMILAB--Pub--96/036-A\\
astro-ph/9602095\\
February 1996 \\
Submitted to {\em Phys.\ Rev.\ D}}
\end{flushright}
\renewcommand{\thefootnote}{\fnsymbol{footnote}}
\vspace{0.15in}
\baselineskip=24pt

\begin{center}
{\Large \bf  Preheating and symmetry restoration \\
 in collisions of vacuum bubbles}\\
\baselineskip=14pt
\vspace{0.75cm}
Edward W.\ Kolb\footnote{Electronic address: {\tt  
rocky@rigoletto.fnal.gov}}\\
{\em NASA/Fermilab Astrophysics Center\\
Fermi National Accelerator Laboratory, Batavia, Illinois~~60510,  
and\\
Department of Astronomy and Astrophysics, Enrico Fermi Institute\\
The University of Chicago, Chicago, Illinois~~ 60637}\\
\vspace{0.4cm}
Antonio Riotto\footnote{Electronic address:
{\tt riotto@fnas01.fnal.gov}}\\
{\em NASA/Fermilab Astrophysics Center\\
Fermi National Accelerator Laboratory, Batavia, Illinois~~60510}\\
\vspace{0.4cm}
\end{center}

\baselineskip=24pt

\begin{quote}
\hspace*{2em} In first-order inflation a phase transition is completed
by the collisions of expanding true-vacuum bubbles.  If bubble
collisions produce large numbers of soft scalar particles carrying
quantum numbers associated with a spontaneously broken symmetry, then
symmetry restoration may occur in a ``pre-heating'' phase in a manner
similar to symmetry restoration in the pre-heating phase of slow-roll
inflation.  Since bubble collisions lead to inhomogeneities, there is
the possibility of inhomogeneous symmetry restoration where
restoration occurs only in the regions of wall collisions.
\vspace*{8pt}

PACS number(s): 98.80.Cq, 11.27.+d

\renewcommand{\thefootnote}{\arabic{footnote}}
\addtocounter{footnote}{-2}
\end{quote}
\end{titlepage}

\newpage

\baselineskip=24pt
\renewcommand{\baselinestretch}{1.5}
\footnotesep=14pt


\def\LHS{{\sc lhs}}
\def\RHS{{\sc rhs}}
\def\GUT{{\sc gut}}
\def\LTE{{\sc lte}}
\def\VEV{{\sc vev}}
\def\beq{\begin{equation}}
\def\eeq{\end{equation}}
\def\beqa{\begin{eqnarray}}
\def\eeqa{\end{eqnarray}}
\def\tr{{\rm tr}}
\def\ph{\tilde{\gamma}}
\def\g{\tilde{g}}
\def\x{{\bf x}}
\def\p{{\bf p}}
\def\k{{\bf k}}
\def\z{{\bf z}}
\def\re#1{{[\ref{#1}]}}
\def\eqr#1{{Eq.\ (\ref{#1})}}
\def\trh{{T_{\rm RH}}}
\def\tph{{T_{\rm PH}}}

To a very good approximation the universe was in local thermodynamic
equilibrium (\LTE) for nearly all of its early development.  However,
there should have been brief, but important, departures from \LTE.
These excursions from equilibrium left an imprint on the universe.
Examples of non-\LTE\ phenomena include baryogenesis, nucleosynthesis,
freeze-out of a massive particle species, decoupling of matter and
radiation, production of topological or non-topological defects in
cosmological phase transitions, inflation, and reheating after
inflation.  In fact, it may be argued that nearly all of
early-universe cosmology is the study of departures from \LTE.  It is
commonly believed that many of the current issues in cosmology require
an understanding of the nontrivial dynamics in the approach to
equilibrium in the early universe.  Nevertheless, despite its immense
relevance, only very recently has substantial effort been devoted to a
detailed understanding of nonequilibrium phenomena in the early
universe.

The non-equilibrium process of interest in this study is the
phenomenon of reheating after inflation.  There are many varieties of
inflation models, but all have an early period of rapid expansion of
the universe where the Robertson--Walker scale factor `accelerates'
(i.e., $\ddot{a}>0$).  At the end of the accelerated-expansion phase
the radiation density of the universe is effectively zero, and the
universe must be `reheated'.\footnote{Of course `re-heated' may be
somewhat of a misnomer since there is no guarantee that the universe
was hot before inflation.}

In `slow-roll' (sometimes referred to as `chaotic') inflation models
\re{linde}, the universe after inflation was dominated by the energy
density contained in the coherent motion of a scalar field known as
the {\it inflaton}, whose potential energy density was responsible for
the accelerated expansion.  Reheating in slow-roll inflation involves
conversion of this coherent scalar-field energy density into into a
thermal distribution of radiation.  In a simple scenario of reheating,
the inflaton field coherently oscillated about the minimum of its
potential until the age of the universe was equal to the lifetime of
the inflaton, then the inflaton decayed, and the decay products
thermalized.

Recent investigations into the non-linear quantum dynamics of scalar
fields have implications for reheating after slow-roll inflation
\re{linde}.  These studies reveal that the scenario by which the
energy density in coherent oscillations of the inflaton field is
converted to radiation may differ significantly from the above
picture, which considered only the linear evolution in time of the
inflation field \re{noneq}.  Quantum nonlinear effects may lead to an
extremely effective dissipational dynamics and explosive particle
production in even the simplest self-interacting theory where single
particle decay is kinematically forbidden. It is possible that almost
all of the energy stored in the form of coherent inflaton oscillations
at the end of inflation is released after only a few oscillation
periods.  The energy is released in the form of inflaton decay
products, whose occupation number is extremely large, and have
energies much smaller than the temperature that would have been
obtained by an instantaneous conversion of the inflaton energy density
into radiation.

Since it requires several scattering times for the low-energy decay
products to form a thermal distribution, it is rather reasonable to
consider the period in which most of the energy density of the
universe was in the form of the non-thermal quanta produced by
inflaton decay as a separate cosmological era.  This is generally
referred to as the `preheating' epoch.

The phenomenon of symmetry restoration during the preheating era has
been investigated recently by Tkachev \re{tkachev} and by Kofman,
Linde, and Starobinski \re{KLSSR} in the framework of typical chaotic
inflationary models.  It was shown that symmetry restoration processes
during the nonequilibrium stage of preheating may be very efficient
with important implications for Grand Unified Theories (\GUT s) and
axions.  Indeed, if a \GUT\ symmetry is restored during the preheating
epoch, the subsequent symmetry breaking phase transition will
reintroduce the problems of monopoles \re{mono} or domain walls
\re{dw}.

In first-order inflation models (generally, any model in which
inflation is completed by a strongly first-order phase transition,
e.g., the extended inflationary scenario proposed by La and
Steinhardt \re{ex}; for a review of first-order inflation models, see
\re{NOBEL}) the universe was dominated by scalar-field vacuum energy
as in slow-roll inflation, but inflation was terminated by the
nucleation of true vacuum bubbles.  At the end of first-order
inflation most of the energy density of the universe was contained in
the bubble walls.  Reheating was instigated by the collisions of
bubble walls, which converted the bubble-wall tension into individual
quanta of the scalar field, which then decayed into normal particles,
which eventually scattered and formed a thermal distribution.

The aim of the present paper is to suggest another situation in which
symmetry restoration can occur efficiently out-of-equilibrium, namely
during the preheating era subsequent to first-order inflation.

As discussed above, the basic idea of reheating in first-order
inflation is essentially the same as in chaotic inflation: energy
initially stored in a coherent scalar field must be converted into
radiation. However, in first-order inflation this releasing of energy
takes place through a number of steps involving both classical and
quantum processes, and a rich phenomenology associated with these
scenarios can arise. For example, it has been suggested that
gravitational waves [\ref{wave1},\ref{wave2}], black holes
[\ref{blackhole1},\ref{blackhole2}] and the baryon asymmetry \re{bau}
may have been produced during the phase transition.  Whether or not
such phenomena actually occur depends in part on the details of
reheating. For instance, in the baryogenesis scenario of Ref.\
\re{bau} it is important to know if the only source of heavy \GUT\
bosons is from primary particles produced in the bubble wall
collisions which, in turn, depends crucially whether the \GUT\
symmetry is restored after bubble collisions, i.e., on the value of
the re-heat temperature, $T_{{\rm RH}}$.\footnote{The reheating
temperature, $\trh$, is usually defined as the temperature of the
universe when the thermal spectrum of radiation was first obtained
after inflation.}

We shall show, however, that similar to what occurs in the chaotic
inflationary scenarios, the details of symmetry restoration may turn
out to be rather independent of $\trh$, and may in fact be quite
complicated, with the symmetry restored in some regions of the
universe, but not others.

In order to keep the discussion as general as possible, we will not
specify any particular first-order inflaton model, but describe the
salient features of the inflaton potential in terms of three
parameters ($\lambda_\sigma$, $\sigma_0$, and $\epsilon$). We denote
the inflaton field by $\sigma$, which has a potential of the general
form suitable to provide for a first-order phase transition. (Table 1
lists the fields and their interactions.) The potential will be
described in terms of a dimensionless coupling constant
$\lambda_\sigma$, a dimensionless constant $\epsilon$ that determines
the splitting between false-vacuum and true-vacuum potential energy
densities, and a mass scale $\sigma_0$, which also plays the role of
the vacuum expectation value when the symmetry is broken.  The mass of
the field will be $\lambda_\sigma^{1/2}\sigma_0$, and the difference
in energy density between the false and true vacuum states will be
denoted as $\Delta V=\epsilon \lambda_\sigma \sigma_0^4$. The
parameter $\epsilon$ must be less than unity for sufficient inflation
to occur.  This also implies that the bubbles of true vacuum formed in
the transition will be ``thin-wall'' bubbles, with wall thickness much
smaller than the radius.

\begin{table}
\caption{\indent Three fields are involved in our consideration:  the
inflaton field $\sigma$; the field $\chi$ into which the domain
walls disperse; and $\phi$, a field whose spontaneously broken
symmetry may be restored by the $\chi$ background.  In some models
$\chi$ and $\sigma$ may be the same field.}
\vspace*{12pt}
\begin{center}
\begin{tabular}{r|l} \hline \hline
Interaction \phantom{XXX}& \phantom{XXX}Potential term \\
\hline
 & \\
inflaton self interaction: \phantom{XXX} &  
 \phantom{XXX} $V_0(\sigma)= \lambda_\sigma(\sigma^2-\sigma_0^2)^2$ 
	 \phantom{XXX}\\
 & \\  \phantom{XXX}
inflaton false-vacuum energy density: \phantom{XXX} &
  \phantom{XXX} $\Delta V = \epsilon \lambda_\sigma \sigma_0^4$ \\
 & \\
$\chi$---$\phi$ interaction: \phantom{XXX} &
 \phantom{XXX}  $V_{\chi\phi}=g\phi^2\chi^2$ \\
 & \\
$\phi$ self interaction: \phantom{XXX} &
 \phantom{XXX}  $V_0(\phi)= \lambda_\phi(\phi^2-\phi_0^2)^2$ \\
 & \\
\hline \hline
\end{tabular}
\end{center}
\end{table}

{}From the few parameters $\lambda_\sigma$, $\epsilon$, and $\sigma_0$,
one can find all the information required about the bubbles formed in
the phase transition.  For instance, in the thin-wall approximation,
the size of a nucleated bubble is given by $R_c\sim
\left(\epsilon\lambda_\sigma^{1/2} \sigma_0\right)^{-1}$.  Bubbles with
a radius smaller than this critical size will not grow, whereas
bubbles larger than the critical size are exponentially disfavored.
Another crucial parameter is the thickness of the wall separating the
true-vacuum region inside from the false-vacuum region outside the
bubble: $\Delta\sim\left(\lambda_\sigma^{1/2}\sigma_0\right)^{-1}$.
The ratio of the bubble-wall thickness to its size is $\Delta/R_C\sim
\epsilon$, which is much less than unity if the thin-wall
approximation is adopted.  Finally, the energy per unit area of the
bubble wall is $\eta\sim \lambda_\sigma^{1/2}\sigma_0^3$.

When a bubble wall forms, false-vacuum energy is transformed into
bubble-wall energy, with the wall energy initially in the form of
static surface energy.  As the bubbles expand converting false vacuum
to true vacuum, more and more of the wall energy becomes kinetic as
the walls become highly relativistic. Numerical simulations
[\ref{wave2},\ref{blackhole1}] demonstrate that during collisions the
walls oscillate through each other, dispersing the kinetic energy at a
rate determined by the frequency of these oscillations. When the
bubbles have slowed after a few oscillations, they then dissipate
their surface energy into particles of typical energy determined by
the wall thickness.

Although the particles produced in the initial collisions of the walls
may play an interesting role in preheating and reheating, in the
following we will concentrate on the implications of the particles
produced by the potential energy density of the bubble walls. Bubble
walls can be envisaged as coherent states of inflaton particles, so
that the typical energy of the products of their decays is simply the
mass of the inflaton. This energy scale is just equal to the inverse
thickness of the wall.

Let's envision the collision of two plane-parallel domain walls.  The
potential energy per unit area of the bubble walls is given by $\eta
\sim \lambda_\sigma^{1/2}\sigma_0^3 $.  Taking the mean energy of the
particles produced in the bubble wall collisions to be of order of the
inverse thickness of the wall, $E\sim\Delta^{-1}$, the mean
number-per-area of particles produced from the potential energy in the
collisions is $ N\simeq \eta/{E}\sim
\lambda_\sigma^{1/2}\sigma_0^3\Delta $.

Let's now assume that the particles are spread out a distance $d$ from
the region of the wall collision.  If we approximate the particle
density as uniform out to a distance $d$, then the particle number
density within the region is simply
\begin{equation}
\label{n}
n=N/d \sim  \lambda_\sigma^{1/2}\sigma_0^3\Delta/d \sim \sigma_0^2/d .
\end{equation}
In the limit that the walls are spherical with radius $R$ and the
collision products instantly fill the bubble interior, then the factor
of $d$ in \eqr{n} should be replaced by $R$.

Eventually the products of bubble-wall collisions will be
redistributed throughout the bubble interior and thermalized. If we
assume that thermalization is instantaneous, the reheating temperature
is found by imposing $\rho_R=(g_*\pi^2/30)T_{{\rm RH}}^4 =\Delta V$,
where $g_*$ is the effective number of degrees of freedom in all the
species of particles formed in the thermalization processes.  Using
$\Delta V = \epsilon \lambda_\sigma \sigma_0^4$ results in a re-heat
temperature of $T_{{\rm RH}}\sim
g_*^{-1/4}\:\epsilon^{1/4}\:\lambda_\sigma^{1/4}\:\sigma_0$.  Let us
now assume that the typical energy of the particles produced through
bubble collisions is smaller than $T_{{\rm RH}}$, i.e.,
$\Delta^{-1}\la T_{{\rm RH}}$, which translates into the condition
(taking $g_*\sim 100$) $\lambda_\sigma\la 10^{-1} \epsilon$.  If this
condition is satisfied, then a period is required for equilibration,
namely for particles to scatter from energies approximately equal to
$\Delta^{-1}$ to a thermal distribution of temperature $T_{{\rm
RH}}$. In addition, since the bubbles were originally empty,
homogenation is not instantaneous, and requires a time at least as
long as the light travel time across a bubble.  If either of these two
time scales is sufficiently long, we may consider the time interval
during which particles do not have a homogeneous thermal distribution
function as a separate epoch: the preheating era.

As a first approximation, during the preheating period the
distribution function of the created particles can be chosen of the
form \re{tkachev}
\begin{equation}
\label{q}
f(\omega)=A\:\delta\left(\omega-E\right),
\end{equation}
where $E=\Delta^{-1}$ and the constant $A$ may be fixed by computing
the number density of particles, $n=(2\pi)^{-3}\int d^3 \!p \, f(p)$,
and setting it equal to the estimate given in \eqr{n}.  Of course, $A$
has mass dimension one.

Let us now imagine that particles $\chi$ are produced in the bubble
wall collisions and are charged under some symmetry group, so
that their mass $m_\chi$ depends upon some scalar field $\phi$ as
$m_\chi^2(\phi) = m_0^2+g\phi^2$.\footnote{Of course by $\phi^2$ and
$\chi^2$ we mean the appropriate sum over the members of the group
representation.}  Here, $g$ represents a combination of numerical
factors and a coupling constant.  As a simple example we might assume
that the $\phi$-dependent mass originates from a potential term of the
form $V_{\chi\phi} = g\phi^2\chi^2$.

As opposed to large-angle scattering processes, forward-scattering
processes do not alter the distribution function of the particles
traversing a gas of quanta, but simply modify the dispersion
relation. This remains true also in the case of a nonequilibrium
system. Forward scattering is manifest, for example, as ensemble
and scalar background corrections to the particle masses.  Since the
forward scattering rate is usually larger than the large-angle
scattering rate responsible for establishing a thermal distribution,
the nonequilibrium ensemble and scalar background corrections are
present even before the initial distribution function, \eqr{q},
relaxes to its thermal value.  These considerations allow us to impose
$\omega^2={\bf p}^2+m_\chi^2(\phi)$ as the dispersion relation for the
particles created by bubble collisions.

We can not use the imaginary-time formalism to determine the effective
potential for the scalar field $\phi$ during the nonequilibrium
preheating period since in the nonequilibrium case there is no
relation between the density matrix of the system and the time
evolution operator, which is of essential importance in the
formalism. There is, however, the real-time formalism of Thermo Field
Dynamics, which suites our purposes \re{tfd}.  The contribution of the
particles created by bubble collisions to the one-loop effective
potential of the scalar field $\phi$ can be written as
\begin{equation}
\Delta V(\phi)= \int\:\frac{d^3  
p}{(2\pi)^3}\:\int_{\infty}^{\omega_{{\bf p}}
(\phi)}\! d\omega \,  f(\omega).
\end{equation}
The first integration in $\omega$ must be done treating $\omega$ as a
free parameter and setting $\omega_{{\bf p}}(\phi) = \sqrt{{\bf
p}^2+m_\chi^2(\phi)}$.  By making use of \eqr{q}, one obtains
\begin{equation}
\Delta V(\phi)=-A\:\int\:\frac{d^3 p}{(2\pi)^3}\:\theta\left[E-
\omega_{{\bf p}}(\phi)\right] \simeq \frac{n}{E}
\left[m_\chi^2(\phi)-E^2\right].
\end{equation}
Since we are interested in the $\phi$-dependent part of the potential,
we can ignore the $nE$ term and the factor of $m_0^2$ in
$m_\chi^2(\phi)$, and write the potential for the non-equilibrium
configuration as $\Delta V(\phi) = B_{{\rm NE}}\phi^2$, where $B_{{\rm
NE}}=gn/E$.  A similar expression was obtained by Tkachev in Ref.\
\re{tkachev}, using the definition of the effective potential as (the
negative of) the pressure of the system, and assuming that the number
of particles does not change on time scales of interest as the field
$\phi$ evolves.

We now use $n=\sigma_0^2/d$ from \eqr{n}, and $E\sim \Delta^{-1} $, to
obtain $B_{{\rm NE}} \sim g \sigma_0^2\Delta /d$.  Of course $d$ will
depend upon the details of the model and the complexities involved in
the completion of the phase transition.  But it is reasonable to
expect, at least initially, that $d$ is of order $\Delta$, so let us
write $d= \xi \Delta$.  Of course as the bosons diffuse into the
bubble interior $\xi$ will change in time, so we expect $\xi$ to grow
and eventually to become much greater than unity.  But initially, at
least, $\xi$ should not be too much larger than unity.  In terms of
$\xi$, we may express $B_{{\rm NE}}$ as $B_{{\rm NE}} \sim g
\sigma_0^2/\xi$.

Now there are two things left to do.  First, we will determine the
conditions under which the non-equilibrium contributions to the
effective potential can restore the symmetry, and then determine the
criterion for the non-equilibrium effects to be more important than
the equilibrium effects obtained after re-heating.

Let us take the $\phi$ tree-level self-interaction potential to be of
the form $V_0(\phi) = \lambda_\phi(\phi^2-\phi_0^2)^2$.  The symmetry
will be restored (i.e., $\phi=0$ will be a stable minimum) if $d^2
V/d\phi^2$ evaluated at $\phi=0 $ is positive, where now $V$ includes
the sum of the tree-level potential and the one-loop correction,
$V=V_0+\Delta V$.  Symmetry restoration will occur due to
non-equilibrium effects if $-\lambda_\phi\phi_0^2 +B_{{\rm NE}} >0$.
This translates into a bound on $\xi$ for symmetry restoration:
\begin{equation}
\label{eq:PREHEAT}
 \frac{g}{\lambda_\phi} \  \frac{\sigma_0^2}{\phi_0^2}  > \xi \ .
\end{equation}
We can imagine three interesting limits depending upon the magnitude  
of the left-hand side (\LHS) of this inequality.   Since we expect  
$\xi$ always to be greater than one, if the \LHS\ is less than unity  
we would expect non-equilibrium effects {\it never} to cause symmetry  
restoration.  If  the  \LHS\ is greater than one but not very large,  
then one might expect temporary restoration of symmetry around the  
regions of bubble collisions.  Then as  $\xi$ starts to grow as the  
bubble interior is filled, the symmetry will be broken when the  
inequality is violated.  Finally, the \LHS\ may be so much greater  
than unity that the symmetry is restored even after the bubble  
interiors are filled.

Of course the symmetry may also be broken after re-heating by thermal
effects.  This can be seen by calculating the $\phi$-dependent term in
the one-loop effective potential obtained by assuming that the system
is in \LTE\ at temperature $\trh$.  Including $V_{\chi\phi} =
g\phi^2\chi^2$, in the high-temperature limit the one-loop thermal
corrections lead to $\Delta V(\phi,T) \sim g T^2 \phi^2 + \lambda_\phi
T^2 \phi^2$.  If we write $\Delta V(\phi,\trh) = B_{{\rm EQ}} \phi^2$,
with $B_{{\rm EQ}} = (g+\lambda_\phi) \trh^2 $, then $B_{{\rm NE}} $
plays a role in non-equilibrium transitions similar to that played by
$B_{{\rm EQ}}$ for thermal transitions.\footnote{Thus, we see that so
far as symmetry restoration is concerned, in the presence of the soft
bosons left behind in the debris of wall collisions, a scalar field
behaves as if it was in \LTE\ at an effective temperature $T_{{\rm
EFF}}^2= B_{{\rm NE}}/(g+\lambda_\phi) \sim \sigma_0^2g
/[\xi(g+\lambda_\phi)]$.}

Symmetry will be restored after re-heating if $-\lambda_\phi\phi_0^2
+B_{{\rm EQ}} >0$, or expressing this as a limit to $\trh$:
$\trh^2>\lambda_\phi\phi_0^2/(g+\lambda_\phi)$.  Now we know $\trh$ in
terms of the parameters of the inflaton potential, so we may express
the criterion for symmetry restoration after re-heating as
\begin{equation}
\label{eq:REHEAT}
 \frac{g+\lambda_\phi}{\lambda_\phi} \  \frac{\sigma_0^2}{\phi_0^2} \  
> \, \sqrt{\frac{g_*}{\epsilon\lambda_\sigma}}\  .
\end{equation}

The condition for symmetry restoration in pre-heating,  
\eqr{eq:PREHEAT}, and the condition
for symmetry restoration in re-heating, \eqr{eq:REHEAT}, are most  
easily contrasted in the limit $g> \lambda_\phi$.  In that limit
\begin{equation}
 \frac{g}{\lambda_\phi} \  \frac{\sigma_0^2}{\phi_0^2}  > \left\{  
\begin{array}{ll} \xi & \hspace*{3em} \mbox{(symmetry restoration  
during pre-heating)} \\
 & \\
\sqrt{g_*/\epsilon\lambda_\sigma} & \hspace*{3em} \mbox{(symmetry  
restoration during re-heating).}
\end{array}
\right. 
\end{equation}

Depending upon the parameters, it is possible to have restoration  
during {\it both} pre-heating and re-heating, during {\it neither}  
pre-heating or re-heating, or during one and not the other.
Of particular interest might be the case where restoration occurs  
only during pre-heating when $\xi$ is not too large.   Then the  
effects of inhomogeneous symmetry restoration will not be erased  
during re-heating.

In conclusion, symmetry restoration may well occur in the preheating
phase following first-order inflation.  Unlike symmetry restoration in
the preheating phase of chaotic inflation, the restoration may be
inhomogeneous after first-order inflation.  The basic point is that
the phase-space density of bosons created in wall collisions is
greatest in regions of wall interactions.  One may imagine situations
where restoration occurs among the debris of wall collisions, but not
in the initially empty interior of the bubbles. In such a case, the
subsequent symmetry breaking restoration might result in creation of
topological defects if the region of wall interactions is large enough
to contain these defects.

Cosmological implications of this possibility require further study.

\vspace{36pt}

\centerline{\bf ACKNOWLEDGMENTS}

EWK and AR are supported by the DOE and NASA under Grant NAG5--2788.

\frenchspacing
\def\prpts#1#2#3{Phys. Reports {\bf #1}, #2 (#3)}
\def\prl#1#2#3{Phys. Rev. Lett. {\bf #1}, #2 (#3)}
\def\prd#1#2#3{Phys. Rev. D {\bf #1}, #2 (#3)}
\def\plb#1#2#3{Phys. Lett. {\bf #1B}, #2 (#3)}
\def\npb#1#2#3{Nucl. Phys. {\bf B#1}, #2 (#3)}
\def\apj#1#2#3{Astrophys. J. {\bf #1}, #2 (#3)}
\def\apjl#1#2#3{Astrophys. J. Lett. {\bf #1}, #2 (#3)}
\begin{picture}(400,50)(0,0)
\put (50,0){\line(350,0){300}}
\end{picture}

\vspace{0.25in}

\def\labelenumi{[\theenumi]}

\begin{enumerate}

\item\label{linde} A. Linde, Phys. Lett. {\bf B219}, 177 (1983);
A. Albrecht and P. J. Steinhardt, Phys. Rev. Lett. {\bf 48}, 1220  
(1982).

\item\label{noneq} D. Boyanovsky, H.J. de Vega, Phys. Rev.
 {\bf D47}, 2343 (1993);
D. Boyanovsky, D.-S. Lee and A. Singh, Phys. Rev. {\bf D48}, 800  
(1993);
D. Boyanovsky, H.J. de Vega and R. Holman, Phys. Rev. {\bf D49}, 2769
 (1994);  and {\it Proceedings of the Second Paris Cosmology
 Colloquium}, Observatoire de Paris, June 1994, p. 125-215, H.J. de  
Vega
 and N. Sanchez Editors, World Scientific, 1995;
L. Kofman, A.D. Linde  and A.A. Starobinsky,
 Phys. Rev. Lett. {\bf 73}, 3195 (1994);
Y. Shtanov,  J. Traschen and R. Brandenberger,
 Phys. Rev. {\bf D51}, 5438 (1995);
A. Dolgov and K. Freese, Phys. Rev. {\bf D51}, 2693 (1995);  
D. Boyanovsky, H.J. de Vega, R. Holman, D.-S. Lee and A. Singh,
 Phys. Rev. {\bf D51}, 4419 (1995);
D. Boyanovsky, M. D'Attanasio, H.J. de  Vega, R. Holman and D.-S. Lee  
and
A. Singh, hep-ph/9505220;
D. Boyanovsky,  M. D'Attanasio, H.J. de Vega, R. Holman and D.-S.  
Lee,
 Phys. Rev. {\bf D52}, 6805 (1995);
L. Kofman, A. Linde and A. Starobinsky, hep-th/9510119;
I. Kaiser, astro-ph/9507108.

\item\label{tkachev} I. Tkachev, OSU-TA-21/95 preprint.

\item\label{KLSSR} L. Kofman, A.D. Linde  and A.A. Starobinsky,
 hep-th/9510119.

\item\label{mono} Ya.B. Zel'dovich and M. Yu. Khlopov,
 Phys. Lett. {\bf B79}, 239 (1978).

\item\label{dw} Ya.B. Zel'dovich, I. Yu. Kobzarev and L. Okun',
 Sov. Phys. JETP {\bf 40}, 1 (1974).

\item\label{ex} D. La and P.J. Steinhardt,
 Phys. Rev. Lett. {\bf 62}, 376 (1989).

\item\label{NOBEL} E. W. Kolb, in {\em The birth and early evolution  
of
our Universe}, Nobel Symposium 79, J. S. Nilsson, B. Gustafsson, and  
B.-S.
Skagerstam, eds.  (World Scientific, Singapore, 1991), p. 199.

\item\label{wave1} M.S. Turner and F. Wilczek,
 Phys. Rev. Lett. {\bf 65}, 3080 (1990).

\item\label{wave2}  R. Watkins and L. Widrow,
 Nucl. Phys. {\bf B374},  446 (1992).

\item\label{blackhole1} S.W. Hawking, I.G. Moss and J.M. Stewart,
 Phys. Rev. {\bf D26}, 2681 (1982).

\item\label{blackhole2} J.D. Barrow, E.J. Copeland, E.W. Kolb and  
A.R. Liddle,
 Phys. Rev. {\bf D43}, 984 (1991).

\item\label{bau}  J.D. Barrow, E.J. Copeland, E.W. Kolb and A.R.  
Liddle,
 Phys. Rev. {\bf D43}, 977 (1991).

\item\label{tfd} H. Umezawa, H. Matsumoto and M. Tachiki,
 {\it Thermo Field Dynamics and Condensed States}, North Holland,  
1982.
See also, P. Elmfors, K. Enqvist and I. Vilja,
 Phys. Lett. {\bf B326},  37 (1994).

\end{enumerate}

\end{document}